\documentclass[prl,twocolumn,amsmath,amssymb,floatfix,reprint,footinbib,superscriptaddress,longbibliography,showkeys]{revtex4-2}
\usepackage[english]{babel}
\usepackage{picinpar,graphicx}
\usepackage{braket}
\usepackage{amsmath}
\usepackage{graphicx}
\usepackage{epstopdf}
\usepackage{dcolumn}
\usepackage{graphicx}
\usepackage{mathrsfs}
\usepackage{mdwlist}
\usepackage{subfigure}
\usepackage{booktabs}
\usepackage{amsmath}
\usepackage{dsfont}
\usepackage{amstext}
\usepackage{amssymb}
\usepackage{amsbsy}
\usepackage{bbm}
\usepackage{amsthm}
\usepackage{graphicx}
\usepackage{textcomp}
\usepackage{color}
\usepackage[colorlinks,citecolor=blue]{hyperref}
\usepackage{lipsum}
\definecolor{Dgreen}{RGB}{0, 100, 0}
\usepackage{url}
\usepackage[colorlinks]{hyperref}
\hypersetup{%
	plainpages=true,
	breaklinks=true,  
	hypertexnames=false,  
	pageanchor=true,
	colorlinks=true,
	linkcolor={blue},
	citecolor={red},
	urlcolor={blue},
	anchorcolor={black}
}

\begin{document}
	
	\title{Spin-Squeezing-Enhanced Charging for Quantum Dicke Batteries}
	\author{Ke-Xiong Yan}
	\affiliation{Fujian Key Laboratory of Quantum Information and Quantum Optics, Fuzhou University, Fuzhou 350116, China}
	\affiliation{Department of Physics, Fuzhou University, Fuzhou, 350116, China}
	
	\author{Jia-Wen Yu}
	\affiliation{Fujian Key Laboratory of Quantum Information and Quantum Optics, Fuzhou University, Fuzhou 350116, China}
	\affiliation{Department of Physics, Fuzhou University, Fuzhou, 350116, China}
	
	\author{Yiming Yu}
	\affiliation{Fujian Key Laboratory of Quantum Information and Quantum Optics, Fuzhou University, Fuzhou 350116, China}
	\affiliation{Department of Physics, Fuzhou University, Fuzhou, 350116, China}
	
	\author{Jun-Hao Lin}
	\affiliation{Fujian Key Laboratory of Quantum Information and Quantum Optics, Fuzhou University, Fuzhou 350116, China}
	\affiliation{Department of Physics, Fuzhou University, Fuzhou, 350116, China}
	
	\author{Shuai Liu}
	\affiliation{School of Physics and Electronic Engineering, Hubei University of Arts and Science, Xiangyang, 441053, China}
	
	\author{Ye-Hong Chen}\thanks{yehong.chen@fzu.edu.cn}
	\affiliation{Fujian Key Laboratory of Quantum Information and Quantum Optics, Fuzhou University, Fuzhou 350116, China}
	\affiliation{Department of Physics, Fuzhou University, Fuzhou, 350116, China}
	\affiliation{Quantum Information Physics Theory Research Team, Center for Quantum Computing, RIKEN, Wako-shi, Saitama 351-0198, Japan}
	
	\author{Yan Xia}\thanks{xia-208@163.com}
	\affiliation{Fujian Key Laboratory of Quantum Information and Quantum Optics, Fuzhou University, Fuzhou 350116, China}
	\affiliation{Department of Physics, Fuzhou University, Fuzhou, 350116, China}
	
	\author{Franco Nori}
	\affiliation{Quantum Information Physics Theory Research Team, Center for Quantum Computing, RIKEN, Wako-shi, Saitama 351-0198, Japan}%
	\affiliation{Department of Physics, University of Michigan, Ann Arbor, Michigan 48109-1040, USA}
	
	\date{\today}
	\begin{abstract}
		High-power Dicke quantum batteries (QBs) typically exploit collective superradiance, whereas intrinsic matter-matter interactions are conventionally considered detrimental. Here, we propose a counterintuitive paradigm: these interactions can be controlled and repurposed as a synergistic resource to enhance charging power and capacity. In the low-excitation limit, transverse interactions induce collective spin squeezing, causing critical mode softening and an exponential enhancement of effective coupling, which significantly boosts charging power. At higher excitations, these interactions act as a macroscopic nonlinear torque. By appropriately aligning this torque, we effectively lower phase-space dynamical barriers, guiding the system along optimal rapid-charging paths. Importantly, this cooperative enhancement remains highly robust under realistic dissipation, outperforming ideal, dissipationless Dicke QBs in specific regimes. Our results provide a blueprint for exploiting matter interactions to design dissipation-resistant, high-performance many-body QBs.
	\end{abstract}
	
	\maketitle
	
\textit{Introduction.}---Quantum thermodynamics seeks to redefine the limits of energy harvesting and storage at the nanoscale, a pursuit that has driven the rapid development of quantum batteries (QBs)~\cite{PhysRevE.79.041129,PhysRevE.87.042123,Kosloff2013,Goold2016,PhysRevLett.122.047702,PhysRevLett.125.180603,Kurizki_Kofman_2022,RevModPhys.96.031001,43n6-rnj3}. By using quantum resources, such as entanglement and collective coherence, these nanoscale devices are predicted to achieve charging performance that surpasses classical limits~\cite{PhysRevLett.118.150601,PhysRevLett.120.117702,Chen2020,PhysRevA.103.052220,jy9l-l8hv}. In particular, collective charging protocols have emerged as a central paradigm, demonstrating that quantum correlations among many subsystems can dramatically accelerate the macroscopic energy transfer process~\cite{Binder2015,PhysRevLett.118.150601,PhysRevLett.120.117702,PhysRevResearch.2.023113,PhysRevLett.125.236402,PhysRevLett.134.130401}.

As a typical platform for realizing such collective advantages, the Dicke model has become a standard framework for many-body QBs~\cite{PhysRev.93.99,Gross1982,PhysRevA.102.023718,PhysRevResearch.6.033181}. By modifying the atomic level structure or the coupling mechanism, such as using three-level atoms~\cite{PhysRevB.109.235432} or two-photon coupling~\cite{PhysRevB.102.245407,Delmonte2021}, the fast-charging capability and the battery capacity of the Dicke model can be further enhanced. However, these extended Dicke models still assume an ideal ensemble of non-interacting atoms and do not take into account the unavoidable direct atom–atom interactions that exist in realistic solid-state or superconducting physics systems~\cite{z8gv-7yyk}.

In the conventional view of QBs, direct matter--matter interactions are often regarded as detrimental, as they are thought to break permutation symmetry and induce local decoherence, thereby suppressing the superradiant charging acceleration enabled by collective coherence~\cite{PhysRevLett.117.073003,Garraway2011,Krantz_2019,Quach2022}. Although recent studies have introduced specific atomic couplings and external drives to tailor collective dynamics~\cite{PhysRevB.105.115405,PhysRevE.107.054125}, they mainly focus on the benefits of external drives while overlooking the energy transfer potential inherent in the interactions themselves. Here, we propose a counterintuitive paradigm: these matter--matter interactions, usually ignored or even eliminated, do not degrade battery performance; instead, with proper control, they become a novel synergistic resource that significantly enhances the charging power of the QB.

By studying the non-equilibrium dynamics across the entire excitation range, we reveal how the direction of matter interactions fundamentally dictates the charging performance of the Dicke battery. In the initial low-excitation stage, $x$-axis interactions in the ferromagnetic regime synergistically exploit spin-squeezing-induced critical mode softening and exponential coupling enhancement to significantly boost charging power. Conversely, $y$- and $z$-axis interactions offer no such advantage in the ferromagnetic regime, only enhancing power in the antiferromagnetic regime. As the battery reaches higher excitations, $x$- and $y$-axis interactions act as non-linear torques~\cite{PhysRevA.47.5138,Luo2025} that directly regulate dynamical barriers in phase space. In the ferromagnetic regime, the $x$-axis interaction lowers this barrier to accelerate charging, while the $y$-axis interaction raises it, suppressing performance; this behavior entirely reverses in the antiferromagnetic regime. The $z$-axis interaction lacks this non-linear torque, and only indirectly modifies the linear drive, yielding an enhancement in the ferromagnetic regime and a suppression in the antiferromagnetic regime. 

Crucially, this interaction-driven enhancement is highly robust. We demonstrate that even under realistic dissipation, within specific parameter regimes, the capacity and charging power of the QB with interaction remain superior to those of an ideal, dissipationless, and interaction-free Dicke battery. This resilience highlights the practical potential of exploiting matter interactions to build dissipation-resistant, high-performance quantum energy storage devices.

\begin{figure}
	\centering
	\includegraphics[scale=0.35]{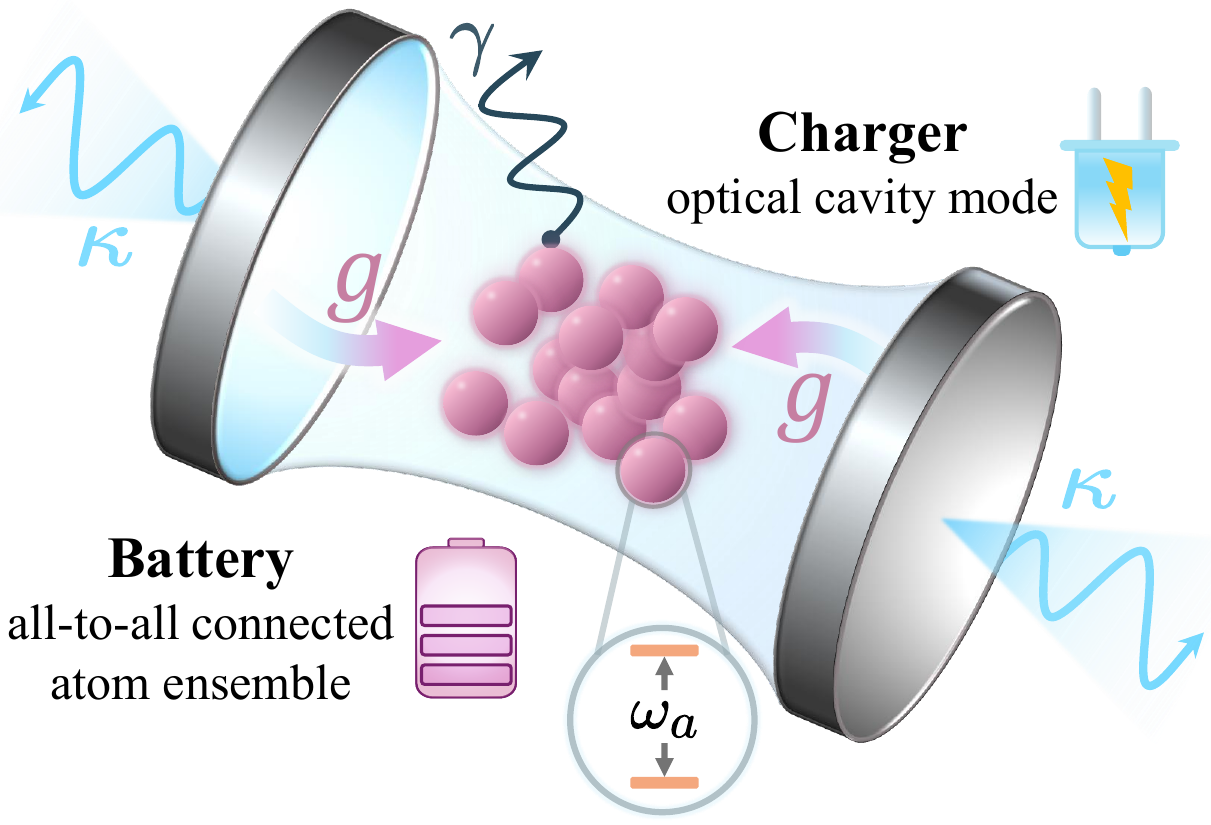}
	\caption{Schematic of the extended Dicke QB. The optical cavity (charger, initial state $|n\rangle$, photon loss $\kappa$) drives an $N$-atom ensemble (battery, initial state $|G\rangle$, atomic decay $\gamma$) via coherent coupling $g$. The inset shows the atomic transition frequency $\omega_a$.}
	\label{fig_model}
\end{figure}
\textit{Model and Indicator.}---We introduce a generalized many-body extension of the Dicke model~\cite{PhysRev.132.2521,Kirton2019,PhysRevResearch.5.033002}, where the interatomic correlation serves as a key resource to generate quantum spin squeezing~\cite{Ma2011}. The total Hamiltonian of the system is given by ($\hbar=1$):
\begin{equation}
	\hat{H} = \omega_a \hat{S}_z + \frac{J_{k}}{N} \hat{S}_{k}^{2}+\omega_c \hat{a}^{\dagger}\hat{a}+\frac{g}{\sqrt{N}}\hat{S}_x (\hat{a}+\hat{a}^\dagger),
	\label{eq1}
\end{equation}
where the first two terms describe the many-body QB $\hat{H}_{B}$, consisting of $N$ two-level atoms with atomic transition frequency $\omega_a$ and collective matter interaction strengths $J_k$ along the $k$-axis ($k \in \{x,y,z\}$). The collective spin components are denoted by $\hat{S}_k = \sum_{i=1}^{N}\sigma_i^k/2$, where $\sigma_i^k$ are the Pauli operators acting on the $i$-th atom. The third term represents the external charger $\hat{H}_{C}$, with $\hat{a}$ ($\hat{a}^{\dagger}$) being the annihilation (creation) operator for the cavity mode with frequency $\omega_c$. The last term $\hat{H}_{\rm int}$ characterizes the collective charging drive with coupling strength $g$, which polarizes the atomic ensemble along the $x$-axis to facilitate energy injection into the battery~\cite{ Marlan1997, PhysRevE.67.066203, Agarwal2012}. A schematic of the model described by Eq.~(\ref{eq1}) is shown in Fig.~\ref{fig_model}. For practical realization of the Hamiltonian $\hat{H}$ in Eq.~(\ref{eq1}), the detailed superconducting circuit implementation and corresponding feasible experimental parameters are discussed in the Supplemental Material\cite{YKX260518}.

We assume that the system is initially in the state $\rho(0) = |G\rangle \langle G| \otimes |n\rangle \langle n|$, where $|G\rangle$ is the ground state of the battery Hamiltonian $\hat{H}_B$, and $|n\rangle$ denotes the $n$-photon Fock state of the charger Hamiltonian $\hat{H}_{C}$. In the absence of dissipation, the unitary evolution of the total system is governed by $\rho(t) = e^{-i\hat{H}t} \rho(0) e^{i\hat{H}t}$. The instantaneous energy stored in the battery at time $t$ is evaluated by the work protocol~\cite{PhysRevLett.122.047702,PhysRevLett.125.180603}:
\begin{equation}
	E(t) = \text{Tr}[\rho(t) \hat{H}_B] - \text{Tr}[\rho(0) \hat{H}_B],
\end{equation}
which quantifies the transient stored energy.  Correspondingly, the instantaneous charging power is defined as the rate of energy transfer, $P(t) = dE(t)/dt$. To capture the peak performance, we evaluate the average charging power~\cite{RevModPhys.96.031001},
\begin{equation}
	P_{\rm avg} = E(t)/t,
\end{equation}
which reaches its maximum value $P_{\rm max}$ at the optimal charging time $t=\tau$, defined as the moment the battery first reaches its energy peak.

\textit{Squeezing-Enhanced Charging.}---The second term in Eq.~(\ref{eq1}), $(J_k/N)\hat{S}_k^2$, represents the one-axis twisting (OAT) interaction~\cite{PhysRevA.47.5138,Ma2011}. This intrinsic nonlinearity induces a state-dependent rotation that shears the quantum uncertainty of an initial coherent spin state into an elongated ellipse on the Bloch sphere, thereby generating spin squeezing~\cite{Ma2011}. The OAT interaction fundamentally reshapes both the ground-state properties and the excitation spectrum. By systematically redistributing the quantum fluctuations, specific OAT orientations can be harnessed to enhance charging power.

In the thermodynamic limit~\cite{PhysRev.58.1098}, we can apply the truncated Holstein-Primakoff (HP) transformation: $\hat{S}_z = \hat{b}^\dagger \hat{b} - N/2$ and $\hat{S}_+ \approx \sqrt{N} \hat{b}^\dagger$ ($\hat{S}_\pm = \hat{S}_x \pm i\hat{S}_y$), provided the atomic excitation number remains macroscopically small compared to the total atom number. This mapping yields three effective Hamiltonians in the HP representation for $k=x,y,z$:
\begin{equation}
	\hat{H}_{k}^{\rm HP}=\Omega_k \hat{b}^\dagger \hat{b} + \hat{H}_{k}^{\rm sq}+\omega_c \hat{a}^\dagger \hat{a} + \frac{g}{2}(\hat{b}+\hat{b}^\dagger)(\hat{a}+\hat{a}^\dagger), \label{eq4}
\end{equation}
where the effective atomic frequencies are $\Omega_z=\omega_a-J_z$, $\Omega_x=\omega_a+J_x /2$ and $\Omega_y=\omega_a+J_y /2$. The structure of the squeezing term $\hat{H}_{k}^{\rm sq}$ depends heavily on the OAT orientation relative to the $z$-axis. For transverse twisting ($k = x, y$), off-diagonal pairing terms emerge, given by $\hat{H}_{x}^{\rm sq}=(J_x/4)(\hat{b}^{\dagger 2}+ \hat{b}^2)$ and $\hat{H}_{y}^{\rm sq}=-(J_y/4)(\hat{b}^{\dagger 2}+ \hat{b}^2)$. These terms describe bosonic squeezing, which serves as the manifestation of collective spin squeezing under this linearized limit~\cite{PhysRevA.68.033821}. Conversely, twisting along the $z$-axis coincides with the direction of the initial spin polarization, preserving the $U(1)$ symmetry of the atomic mode. This configuration results in a pure scalar frequency shift ($\hat{H}_{z}^{\rm sq}=0$) without generating any squeezing.

To diagonalize the atomic sector, we perform a Bogoliubov transformation: $\hat{b} = \hat{\beta} \cosh r_k + \hat{\beta}^\dagger \sinh r_k$, where the squeezing parameters $r_k$ are determined by $\tanh(2r_x)=-J_x/(J_x+2\omega_a)$ and $\tanh(2r_y)=J_y/(J_y+2\omega_a)$. In the squeezed quasiparticle frame, the effective Hamiltonians $\hat{H}_x^{\rm HP}$ and $\hat{H}_{y}^{\rm HP}$ take the same form
\begin{equation}
	\hat{H}_k =\Delta_k \hat{\beta}^\dagger \hat{\beta} + \omega_c \hat{a}^\dagger \hat{a} + \frac{g e^{r_{k}}}{2} (\hat{\beta} + \hat{\beta}^\dagger)(\hat{a} + \hat{a}^\dagger), \label{eq5}
\end{equation}
where $k=x,y$ and $\Delta_k = \sqrt{\omega_a(\omega_a+J_k)}$ is the renormalized atomic frequency. Notably, the effective light-matter coupling is scaled by a factor of $e^{r_k}/2$. Therefore, tuning the interactions to increase $r_k $ leads to an exponential enhancement of this coupling~\cite{PhysRevLett.120.093601,Qin2024}.

Diagonalizing Eq.~(\ref{eq5}) yields the normal mode frequencies $\omega_{\pm}=\sqrt{\Delta_{k}^{2}\pm g\Delta_{k}e^{r_{k}}}$. As the interaction drives the lower mode frequency toward zero ($\omega_{-}\rightarrow0$), the system exhibits soft-mode criticality. In this regime, based on the exact analytical solutions for QBs~\cite{PhysRevB.98.205423}, the dynamics are primarily governed by this soft mode. The transient stored energy asymptotically scales as
\begin{equation}
	E(t)\propto\Delta_{k}e^{2r_{k}}\left(\frac{\sin(t\omega_{-})}{\omega_{-}}\right)^{2}.
\end{equation}
The average charging power becomes~\cite{YKX260518}
\begin{equation}
	P_{\mathrm{avg}}(t)\propto\frac{\Delta_{k}e^{2r_{k}}}{\omega_{-}}\left[\frac{\sin^{2}(t\omega_{-})}{(t\omega_{-})}\right].
\end{equation}
The global maximum of the function $f(x)=\sin^{2}(x)/x$ is a dimensionless constant. By substituting the explicit form of $\omega_{-}$, we directly obtain the unified asymptotic scaling formula for the maximum average charging power:
\begin{equation}
	P_{\mathrm{max}}^{(k)}\propto\frac{e^{2r_{k}}}{\sqrt{1-ge^{r_{k}}/\Delta_{k}}}.
\end{equation}

This unified expression directly links the charging performance to the soft-mode criticality of the system, which exhibits distinct behaviors across different interaction regimes. In the ferromagnetic regime ($J_k <0$), the effects of mode softening and coupling enhancement strongly depend on the interaction axis. For the direction-matched $x$-axis interaction, collective spin squeezing amplifies the effective coupling, while the energy gap $\Delta_x$ simultaneously narrows (mode softening). These two effects are mutually synergistic, driving the denominator toward zero and triggering a divergent peak power. Conversely, for an attractive $y$-axis interaction, the effects are mutually antagonistic: although the energy gap $\Delta_y$ decreases, the lack of appropriate squeezing sharply suppresses the effective coupling, fundamentally decreasing the overall charging power. 

In the antiferromagnetic regime ($J_k >0$), the impacts of the $x$- and $y$-axes are entirely reversed. The $x$-axis interaction now suppresses the coupling and widens the energy gap, severely restricting the energy transfer. The repulsive $y$-axis interaction restores squeezing and yields an exponentially enhanced coupling. However, this is accompanied by an increased energy gap $\Delta_y$. These mutually antagonistic effects prevent the denominator from vanishing, allowing only a moderate power enhancement. Finally, the $z$-axis interaction ($r_z = 0$) generates no spin squeezing in either regime and purely modifies the effective detuning $\Delta_z$. Consequently, only the $x$-axis configuration in the ferromagnetic regime can fully exploit the soft-mode criticality to maximize energy transfer.

\begin{figure}
	\centering
	\includegraphics[width=\linewidth]{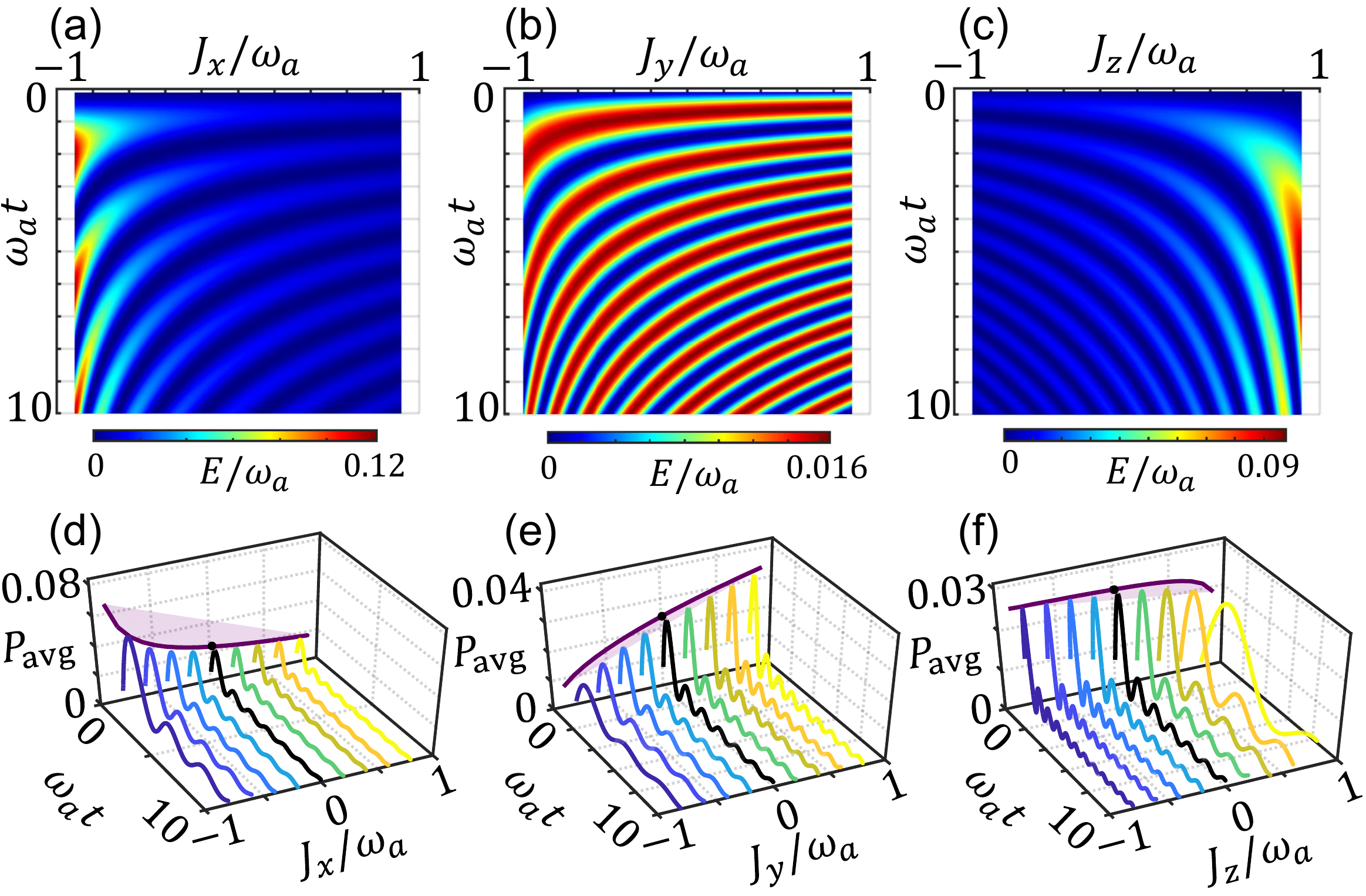}
	\caption{HP-regime charging dynamics. (a,b,c) Stored energy $E/\omega_a$ and (d,e,f) average power $P_{\rm avg}$ vs interaction strength $J_k/\omega_a$ and time $\omega_a t$ for $k = x, y, z$ interactions, respectively. Solid purple curves trace the peak power. The system is initialized in $|G\rangle \otimes |n=5\rangle$ with coupling $g = 0.08\omega_a$. The cavity is tuned to resonance ($\omega_c = \Delta_k$).}
	\label{fig1}
\end{figure}
These theoretical predictions are numerically confirmed in Fig.~\ref{fig1}, highlighting a strong directional and regime dependence. In the ferromagnetic regime, as the system approaches the critical region ($J_x \to -\omega_a$), the $x$-axis interaction [Figs.~\ref{fig1}(a,d)] exploits the synergistic enhancement to simultaneously reduce the charging time and boost the maximum stored energy to $\sim 0.12\omega_a$, yielding a peak power of $P_{\rm avg} \sim 0.08\omega_a^2$. In contrast, the ferromagnetic $y$-axis interaction [Figs.~\ref{fig1}(b,e)] selectively traps the system in a low-capacity state ($\sim 0.016\omega_a$), which inherently constrains its absolute power output despite faster Rabi oscillations~\cite{Ashhab2006,PhysRevB.78.054512,PhysRevB.80.212503,PhysRevA.92.063830,PhysRevLett.126.023602,PhysRevLett.133.033603}. The $z$-axis interaction [Figs.~\ref{fig1}(c,f)] generates no spin squeezing and purely modifies the effective detuning, yielding only a moderate enhancement in energy transfer.

\textit{Nonlinearity-Enhanced Charging.}---While squeezing effectively captures the initial acceleration, the linear nature of the Holstein-Primakoff approximation intrinsically fails to fully describe the charging cycle as the battery enters the high-excitation regime ($\langle \hat{S}_z \rangle \sim N/2$)~\cite{PhysRevE.67.066203}. To evaluate the complete charging dynamics, the collective state of the atomic ensemble is mapped onto a semiclassical phase space, parameterized by the polar angle $\theta \in [0, \pi]$ and the azimuthal phase $\phi \in [0, 2\pi]$ on the Bloch sphere~\cite{PhysRevA.6.2211,Ma2011}.

 We define the normalized energy inversion as $w = \langle \hat{S}_z \rangle / S \equiv \cos\theta$, where $S = N/2$ is the total collective spin length. Here, $w = -1$ ($\theta = \pi$) corresponds to the fully depleted ground state, and $w = 1$ ($\theta = 0$) represents the fully charged state.

By applying the Hamiltonian equations of motion to the semiclassical energy functional derived from Eq.~(\ref{eq1}) and adiabatically eliminating the cavity field~\cite{Marlan1997,PhysRevE.67.066203,Agarwal2012}, we obtain the macroscopic nonlinear charging rate $\dot{w}$~\cite{YKX260518}:
\begin{equation}
	\dot{w} = 
	\begin{cases} 
		\Omega v + J_x u v & \text{for } k = x, \\
		\Omega v - J_y u v & \text{for } k = y, \\
		\Omega v & \text{for } k = z,
	\end{cases}
	\label{eq:dotw}
\end{equation}
where $u = \sin\theta\cos\phi$ and $v = \sin\theta\sin\phi$ are the normalized transverse polarizations. Assuming the charger is initialized in a macroscopic Fock state ($n \gg 1$), the vanishing relative photon fluctuations allow the cavity field to dynamically establish a well-defined phase and act as a steady classical drive $\Omega \propto g\sqrt{n/N}$~\cite{Marlan1997,Gross1982}. Consequently, the term $\Omega v$ in the equations of motion dictates the power injected by this macroscopic charger, whereas the components proportional to $J_k$ act as an internal nonlinear torque.

\begin{figure}
	\centering
	\includegraphics[width=\linewidth]{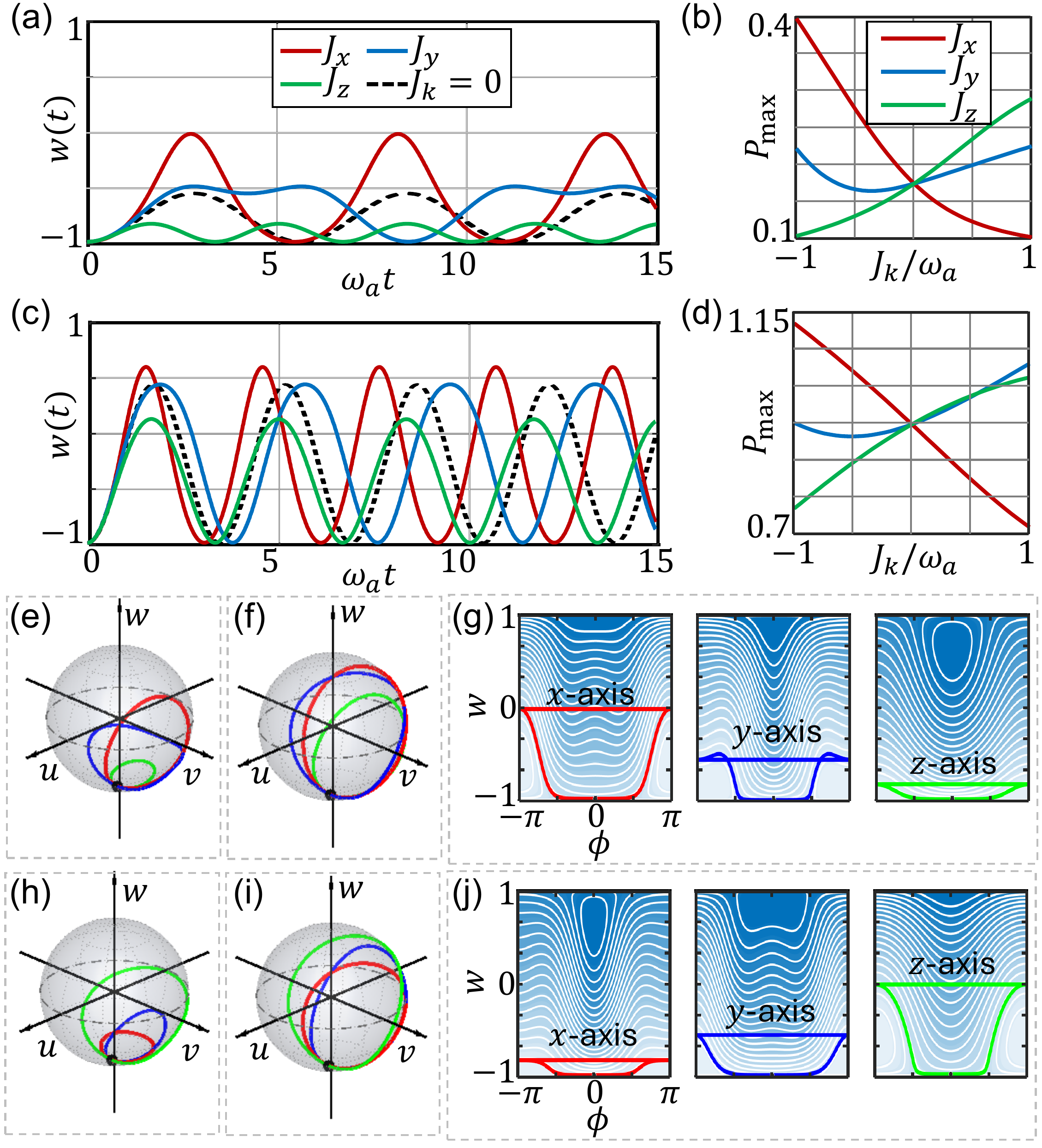}
	\caption{Mean-field charging dynamics. Panels (a,b,e,h) use a moderate drive ($\Omega=0.5\omega_a$), while (c,d,f,i) use a strong drive ($\Omega=1.5\omega_a$). (a,c) Energy inversion $w(t)$, compared to the ideal non-interacting benchmark ($J=0$, black dashed). (b,d) Average power $P_{\rm avg}$ vs $J_k/\omega_a$. (e,f) and (h,i) show Bloch sphere trajectories for $J_k = -0.9\omega_a$ and $0.9\omega_a$, respectively. (g,j) Semiclassical energy landscapes for $J_k = -0.9\omega_a$ and $0.9\omega_a$ ($\Omega=0.5\omega_a$) with overlaid projected trajectories (solid curves).}
	\label{fig2}
\end{figure}
The charging efficiency in the high-excitation regime is fundamentally dictated by how internal interactions act as a macroscopic nonlinear torque to regulate the dynamical barriers in phase space. Under a moderate drive [Figs.~\ref{fig2}(a,b,e,h)], this torque strongly depends on both the interaction axis and the interaction regime.

In the ferromagnetic regime ($J_k < 0$), a direction-matched $x$-axis interaction cooperates synergistically with the driving cavity field. This macroscopic torque effectively lowers the dynamical barrier in the phase-space energy landscape [Fig.~\ref{fig2}(g), $x$-axis], significantly extending the Bloch sphere trajectory [Fig.~\ref{fig2}(e)] and yielding a peak energy storage ($w \approx 0.5$) that distinctly outperforms the non-interacting benchmark ($J=0$). Conversely, the $y$-axis interaction acts antagonistically in this regime; it directly opposes the drive by creating a severe transverse barrier [Fig.~\ref{fig2}(g), $y$-axis] that traps the system in low-amplitude oscillations. Similarly, the $z$-axis interaction lacks the ability to produce a beneficial nonlinear torque, merely inducing a dynamic detuning that shifts the system out of resonance and severely suppresses the energy exchange.

In the antiferromagnetic regime ($J_k>0$), the physical impact of the transverse interactions are entirely reversed. As visually confirmed by the semiclassical energy landscapes in Fig.~\ref{fig2}(j), the $x$-axis interaction now raises the dynamical barrier, severely restricting the Bloch sphere trajectory [Fig.~\ref{fig2}(h)] and becoming highly detrimental to the charging process. Instead, it is the $y$-axis interaction that lowers the dynamical barrier in this regime, providing a cooperative torque that enhances the charging trajectory relative to the $x$-axis. The $z$-axis interaction remains unhelpful for barrier regulation.

These torque-driven, direction-dependent effects are clearly reflected in the average charging power $P_{\rm avg}$ [Figs.~\ref{fig2}(b,d)]. Crucially, the capacity to harness interactions as a synergistic resource is highly conditional: the $x$-axis interaction yields a substantial power enhancement in the ferromagnetic regime, whereas the $y$-axis interaction takes the advantageous role in the antiferromagnetic regime. Notably, even when the drive amplitude is substantially increased ($\Omega = 1.5\omega_a$) to overpower the internal matter interactions [Figs.~\ref{fig2}(c,d,f,i)], this fundamental divergence in phase-space trajectories and charging performance persists. Ultimately, these results demonstrate that appropriately aligning the nonlinear torque with the specific interaction regime is the essential key to lowering dynamical barriers and achieving superior charging performance.

\textit{Effects of Dissipation.}---In realistic experimental platforms, such as superconducting circuits~\cite{Hu2022} or inorganic and organic nanostructures exhibiting collective effects~\cite{Quach2022,Camposeo2025}, QBs are inevitably coupled to their noisy environments. To evaluate the practical viability of the interaction-induced enhancement, we incorporate the dominant dissipative channels into the macroscopic semiclassical dynamics, specifically the cavity photon loss rate $\kappa$ and the atomic spontaneous emission rate $\gamma$. In the macroscopic leaky cavity limit ($n \gg N$), cavity dissipation causes the effective classical driving amplitude to decay exponentially as $\Omega(t) = \Omega_0 e^{-\kappa t/2}$, where $\Omega_0$ is the initial drive amplitude~\cite{Carmichael1993,Walls2025}. Concurrently, atomic spontaneous emission introduces linear relaxation and dephasing. To incorporate these dissipative effects, we employ the Heisenberg-Langevin equation for any system operator $\hat{O}$~\cite{PhysRevA.100.022115}:
\begin{equation}
	\frac{d\langle\hat{O}\rangle}{dt}=i\langle[\hat{H}_{\mathrm{eff}},\hat{O}]\rangle+\langle\mathcal{L}^{\dagger}(\hat{O})\rangle,
\end{equation}
where the adjoint Lindblad superoperator $\mathcal{L}^{\dagger}(\hat{O})$ describing the independent local atomic decay at a rate $\gamma$ is explicitly given by
\begin{equation}
	\mathcal{L}^{\dagger}(\hat{O})=\gamma\sum_{i=1}^{N}\left(\sigma_{+}^{(i)}\hat{O}\sigma_{-}^{(i)}-\frac{1}{2}\{\sigma_{+}^{(i)}\sigma_{-}^{(i)},\hat{O}\}\right).
\end{equation}
Under the semiclassical mean-field approximation, the dissipative equations of motion for the $x$-axis interaction become~\cite{YKX260518}:
\begin{equation}
	\begin{cases}
		\dot{u}=-\omega_{a}v-\frac{\gamma}{2}u, \\ 
		\dot{v}=\omega_{a}u-J_{x}uw-\Omega_{0}e^{-\kappa t/2}w-\frac{\gamma}{2}v, \\ 
		\dot{w}=J_{x}uv+\Omega_{0}e^{-\kappa t/2}v-\gamma(w+1).
	\end{cases}
\end{equation}

\begin{figure}
	\centering
	\includegraphics[width=\linewidth]{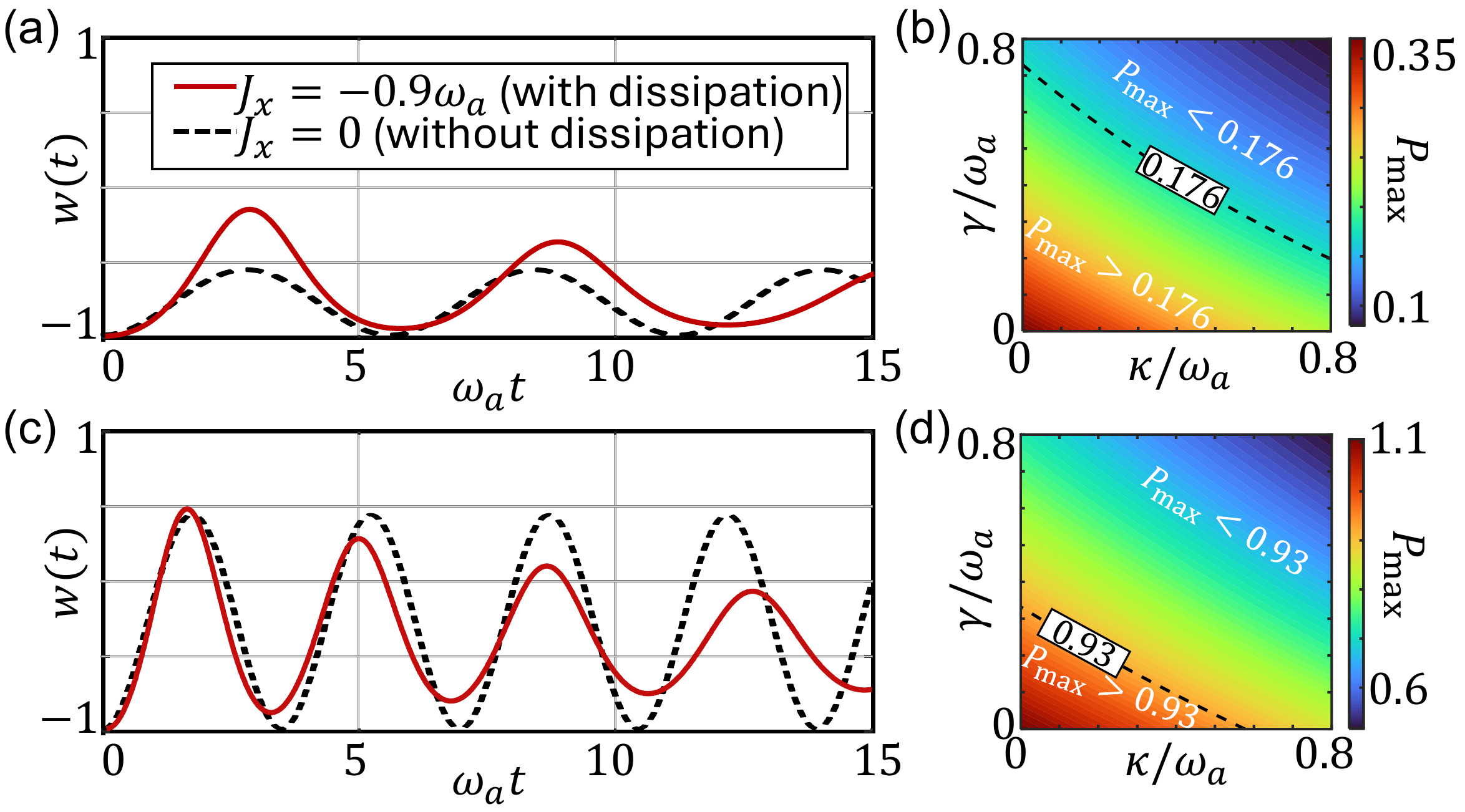}
	\caption{Robustness against dissipation. (a,c) Energy inversion $w(t)$ for $\Omega=0.5\omega_a$ and $1.5\omega_a$, comparing the dissipative $x$-axis interacting battery (red solid, $J_x=-0.9\omega_a, \kappa=\gamma=0.1\omega_a$) to the ideal non-interacting benchmark (black dashed, $J_x=\kappa=\gamma=0$). (b,d) Maximum average power $P_{\rm max}$ in the $\kappa$--$\gamma$ space. Black dashed contours denote the ideal benchmark limits ($P_{\rm max}=0.176\omega_a^2$ and $0.93\omega_a^2$), below which the dissipative interacting battery retains superiority.}
	\label{fig3}
\end{figure}

Figures~\ref{fig3}(a,c) present the dissipative charging dynamics $w(t)$ under a moderate decay rate ($\kappa = \gamma = 0.1\omega_a$) for both moderate ($\Omega_0 = 0.5\omega_a$) and strong ($\Omega_0 = 1.5\omega_a$) drives. Remarkably, despite active energy dissipation and decoherence, the direction-matched $x$-axis interacting battery ($J_x = -0.9\omega_a$) exhibits an accelerated charging process that distinctly surpasses the peak capacity of an ideal, dissipationless Dicke battery ($J_x = 0$) during early cycles. This transient advantage is powered by the constructive nonlinear torque $J_x uw$, which acts as a macroscopic booster to rapidly pump energy into the system before dissipation takes over.                                                                                   

Figures~\ref{fig3}(b,d) map the maximum average power $P_{\text{max}}$ in the $\kappa$-$\gamma$ parameter space. The black dashed contours mark the performance thresholds of the ideal, interaction-free limits ($P_{\text{max}} = 0.176\omega_a^2$ and $0.93\omega_a^2$, respectively). The expansive parameter regions below these contours confirm that the $x$-axis configuration successfully resists environmental noise, consistently outperforming the ideal Dicke battery benchmark across a wide range of loss rates.

\textit{Conclusion.}---We have demonstrated that properly aligning interatomic interaction with the operational regime systematically enhances the charging power and capacity of the QB across all excitation ranges. In the low-excitation limit, the direction-matched configuration leverages spin-squeezing criticality to induce mode softening and accelerate the initial energy transfer. As the QB enters the high-excitation regime, this interaction manifests as a macroscopic nonlinear torque that effectively lowers phase-space dynamical barriers to sustain rapid charging.

This cooperative enhancement remains highly robust against realistic environmental noise, including cavity photon loss and atomic spontaneous emission. Under these open-system conditions, the interacting battery consistently outperforms the ideal, interaction-free Dicke benchmark in both capacity and charging power. This proven resilience provides a practical blueprint for engineering high-performance quantum energy storage devices.

\textit{Acknowledgments.}---Y.-H.C. was supported by the National Natural Science Foundation of
China (Grants No. 12304390 and No. 12574386), the National Postdoctoral Overseas Talent Recruitment Program of China, the Fujian 100 Talents Program, and the Fujian Minjiang Scholar Program. Y.X. was supported by the National Natural Science Foundation of China under Grant No. 62471143, the Key Program of National Natural Science Foundation of Fujian Province under Grant No. 2024J02008. F.N. is supported in part by the Japan Science and Technology Agency (JST) [via the CREST Quantum Frontiers program Grant No. JPMJCR24I2, the Quantum Leap Flagship Program (Q-LEAP), the Moonshot R$\&$D Grant Number JPMJMS256E, and the ASPIRE program (Grant Number JPMJAP2513)].

\bibliography{reference}

@article{43n6-rnj3,
	title = {Giant-Atom Quantum Batteries: Lossless Energy Transfer via Interference Engineering},
	author = {Yan, Ke-Xiong and Liu, Yang and Xiao, Yang and Lin, Jun-Hao and Song, Jie and Chen, Ye-Hong and Nori, Franco and Xia, Yan},
	journal = {Phys. Rev. Lett.},
	volume = {136},
	issue = {18},
	pages = {180401},
	numpages = {7},
	year = {2026},
	month = {May},
	publisher = {American Physical Society},
	doi = {10.1103/43n6-rnj3},
	url = {https://link.aps.org/doi/10.1103/43n6-rnj3}
}

@book{Kurizki_Kofman_2022, place={Cambridge}, title={Thermodynamics and Control of Open Quantum Systems}, publisher={Cambridge University Press}, author={Kurizki, Gershon and Kofman, Abraham G.}, year={2022}}

@article{Binder2015,
	title = {Quantacell: powerful charging of quantum batteries},
	volume = {17},
	ISSN = {1367-2630},
	url = {http://dx.doi.org/10.1088/1367-2630/17/7/075015},
	DOI = {10.1088/1367-2630/17/7/075015},
	number = {7},
	journal = {New J. Phys.},
	publisher = {IOP Publishing},
	author = {Binder,  Felix C and Vinjanampathy,  Sai and Modi,  Kavan and Goold,  John},
	year = {2015},
	month = {July},
	pages = {075015}
}

@article{PhysRevLett.118.150601,
	title = {Enhancing the Charging Power of Quantum Batteries},
	author = {Campaioli, Francesco and Pollock, Felix A. and Binder, Felix C. and C\'eleri, Lucas and Goold, John and Vinjanampathy, Sai and Modi, Kavan},
	journal = {Phys. Rev. Lett.},
	volume = {118},
	issue = {15},
	pages = {150601},
	numpages = {6},
	year = {2017},
	month = {Apr},
	publisher = {American Physical Society},
	doi = {10.1103/PhysRevLett.118.150601},
	url = {https://link.aps.org/doi/10.1103/PhysRevLett.118.150601}
}

@article{PhysRevLett.120.117702,
	title = {High-Power Collective Charging of a Solid-State Quantum Battery},
	author = {Ferraro, Dario and Campisi, Michele and Andolina, Gian Marcello and Pellegrini, Vittorio and Polini, Marco},
	journal = {Phys. Rev. Lett.},
	volume = {120},
	issue = {11},
	pages = {117702},
	numpages = {6},
	year = {2018},
	month = {Mar},
	publisher = {American Physical Society},
	doi = {10.1103/PhysRevLett.120.117702},
	url = {https://link.aps.org/doi/10.1103/PhysRevLett.120.117702}
}

@article{PhysRevLett.122.047702,
	title = {Extractable Work, the Role of Correlations, and Asymptotic Freedom in Quantum Batteries},
	author = {Andolina, Gian Marcello and Keck, Maximilian and Mari, Andrea and Campisi, Michele and Giovannetti, Vittorio and Polini, Marco},
	journal = {Phys. Rev. Lett.},
	volume = {122},
	issue = {4},
	pages = {047702},
	numpages = {5},
	year = {2019},
	month = {Feb},
	publisher = {American Physical Society},
	doi = {10.1103/PhysRevLett.122.047702},
	url = {https://link.aps.org/doi/10.1103/PhysRevLett.122.047702}
}

@article{PhysRevLett.125.180603,
	title = {Quantum Coherence and Ergotropy},
	author = {Francica, G. and Binder, F. C. and Guarnieri, G. and Mitchison, M. T. and Goold, J. and Plastina, F.},
	journal = {Phys. Rev. Lett.},
	volume = {125},
	issue = {18},
	pages = {180603},
	numpages = {8},
	year = {2020},
	month = {Oct},
	publisher = {American Physical Society},
	doi = {10.1103/PhysRevLett.125.180603},
	url = {https://link.aps.org/doi/10.1103/PhysRevLett.125.180603}
}

@article{PhysRevResearch.2.023113,
	title = {Bounds on the capacity and power of quantum batteries},
	author = {Juli\`a-Farr\'e, Sergi and Salamon, Tymoteusz and Riera, Arnau and Bera, Manabendra N. and Lewenstein, Maciej},
	journal = {Phys. Rev. Res.},
	volume = {2},
	issue = {2},
	pages = {023113},
	numpages = {16},
	year = {2020},
	month = {May},
	publisher = {American Physical Society},
	doi = {10.1103/PhysRevResearch.2.023113},
	url = {https://link.aps.org/doi/10.1103/PhysRevResearch.2.023113}
}

@article{PhysRevLett.125.236402,
	title = {Quantum Advantage in the Charging Process of {Sachdev-Ye-Kitaev} Batteries},
	author = {Rossini, Davide and Andolina, Gian Marcello and Rosa, Dario and Carrega, Matteo and Polini, Marco},
	journal = {Phys. Rev. Lett.},
	volume = {125},
	issue = {23},
	pages = {236402},
	numpages = {6},
	year = {2020},
	month = {Dec},
	publisher = {American Physical Society},
	doi = {10.1103/PhysRevLett.125.236402},
	url = {https://link.aps.org/doi/10.1103/PhysRevLett.125.236402}
}

@article{PhysRevLett.134.130401,
	title = {Large Collective Power Enhancement in Dissipative Charging of a Quantum Battery},
	author = {Pokhrel, Sagar and Gea-Banacloche, Julio},
	journal = {Phys. Rev. Lett.},
	volume = {134},
	issue = {13},
	pages = {130401},
	numpages = {6},
	year = {2025},
	month = {Mar},
	publisher = {American Physical Society},
	doi = {10.1103/PhysRevLett.134.130401},
	url = {https://link.aps.org/doi/10.1103/PhysRevLett.134.130401}
}

@article{PhysRevE.107.054125,
	title = {Quantum battery based on dipole-dipole interaction and external driving field},
	author = {Zhang, Wuji and Wang, Shuyue and Wu, Chunfeng and Wang, Gangcheng},
	journal = {Phys. Rev. E},
	volume = {107},
	issue = {5},
	pages = {054125},
	numpages = {10},
	year = {2023},
	month = {May},
	publisher = {American Physical Society},
	doi = {10.1103/PhysRevE.107.054125},
	url = {https://link.aps.org/doi/10.1103/PhysRevE.107.054125}
}

@article{Quach2022,
	title = {Superabsorption in an organic microcavity: Toward a quantum battery},
	volume = {8},
	ISSN = {2375-2548},
	url = {http://dx.doi.org/10.1126/sciadv.abk3160},
	DOI = {10.1126/sciadv.abk3160},
	number = {2},
	journal = {Sci. Adv.},
	publisher = {American Association for the Advancement of Science (AAAS)},
	author = {Quach,  James Q. and McGhee,  Kirsty E. and Ganzer,  Lucia and Rouse,  Dominic M. and Lovett,  Brendon W. and Gauger,  Erik M. and Keeling,  Jonathan and Cerullo,  Giulio and Lidzey,  David G. and Virgili,  Tersilla},
	year = {2022},
	month = {Jan} 
}

@article{Krantz_2019,
	title = {A quantum engineer’s guide to superconducting qubits},
	author = {Krantz, P. and Kjaergaard, M. and Yan, F. and Orlando, T. P. and Gustavsson, S. and Oliver, W. D.},
	journal = {Appl. Phys. Rev.},
	volume = {6},
	issue = {2},
	pages = {},
	numpages = {},
	year = {2019},
	month = {June},
	publisher = {AIP Publishing},
	doi = {10.1063/1.5089550},
	url = {http://dx.doi.org/10.1063/1.5089550}
}

@article{RevModPhys.96.031001,
	title = {Colloquium: Quantum batteries},
	author = {Campaioli, Francesco and Gherardini, Stefano and Quach, James Q. and Polini, Marco and Andolina, Gian Marcello},
	journal = {Rev. Mod. Phys.},
	volume = {96},
	issue = {3},
	pages = {031001},
	numpages = {30},
	year = {2024},
	month = {Jul},
	publisher = {American Physical Society},
	doi = {10.1103/RevModPhys.96.031001},
	url = {https://link.aps.org/doi/10.1103/RevModPhys.96.031001}
}

@article{Luo2025,
	title = {Hamiltonian engineering of collective {XYZ} spin models in an optical cavity},
	volume = {21},
	ISSN = {1745-2481},
	url = {http://dx.doi.org/10.1038/s41567-025-02866-0},
	DOI = {10.1038/s41567-025-02866-0},
	number = {6},
	journal = {Nat. Phys.},
	publisher = {Springer Science and Business Media LLC},
	author = {Luo,  Chengyi and Zhang,  Haoqing and Chu,  Anjun and Maruko,  Chitose and Rey,  Ana Maria and Thompson,  James K.},
	year = {2025},
	month = Apr,
	pages = {916–923}
}

@article{PhysRevB.102.245407,
	title = {Ultrafast charging in a two-photon {Dicke} quantum battery},
	author = {Crescente, Alba and Carrega, Matteo and Sassetti, Maura and Ferraro, Dario},
	journal = {Phys. Rev. B},
	volume = {102},
	issue = {24},
	pages = {245407},
	numpages = {12},
	year = {2020},
	month = {Dec},
	publisher = {American Physical Society},
	doi = {10.1103/PhysRevB.102.245407},
	url = {https://link.aps.org/doi/10.1103/PhysRevB.102.245407}
}

@article{Delmonte2021,
	title = {Characterization of a Two-Photon Quantum Battery: Initial Conditions,  Stability and Work Extraction},
	volume = {23},
	ISSN = {1099-4300},
	url = {http://dx.doi.org/10.3390/e23050612},
	DOI = {10.3390/e23050612},
	number = {5},
	journal = {Entropy},
	publisher = {MDPI AG},
	author = {Delmonte,  Anna and Crescente,  Alba and Carrega,  Matteo and Ferraro,  Dario and Sassetti,  Maura},
	year = {2021},
	month = May,
	pages = {612}
}

@article{Chen2020,
	title = {Charging Quantum Batteries with a General Harmonic Driving Field},
	volume = {532},
	ISSN = {1521-3889},
	url = {http://dx.doi.org/10.1002/andp.201900487},
	
	number = {4},
	journal = {Ann. Phys.},
	publisher = {Wiley},
	author = {Chen,  Jie and Zhan,  Liyao and Shao,  Lei and Zhang,  Xingyu and Zhang,  Yuyu and Wang,  Xiaoguang},
	year = {2020},
	month = mar 
}

@article{PhysRevA.103.052220,
	title = {Lower and upper bounds of quantum battery power in multiple central spin systems},
	author = {Peng, Li and He, Wen-Bin and Chesi, Stefano and Lin, Hai-Qing and Guan, Xi-Wen},
	journal = {Phys. Rev. A},
	volume = {103},
	issue = {5},
	pages = {052220},
	numpages = {9},
	year = {2021},
	month = {May},
	publisher = {American Physical Society},
	doi = {10.1103/PhysRevA.103.052220},
	url = {https://link.aps.org/doi/10.1103/PhysRevA.103.052220}
}

@article{PhysRevB.105.115405,
	title = {Extended {D}icke quantum battery with interatomic interactions and driving field},
	author = {Dou, F.-Q. and Lu, Y.-Q. and Wang, Y.-J. and Sun, J.-A.},
	journal = {Phys. Rev. B},
	volume = {105},
	issue = {11},
	pages = {115405},
	numpages = {13},
	year = {2022},
	month = {Mar},
	publisher = {American Physical Society},
	doi = {10.1103/PhysRevB.105.115405},
	url = {https://link.aps.org/doi/10.1103/PhysRevB.105.115405}
}

@article{jy9l-l8hv,
	title = {Efficient charging of multiple open quantum batteries through dissipation and pumping},
	author = {Dias, Josephine and Wang, Hui and Nemoto, Kae and Nori, Franco and Munro, William J.},
	journal = {Phys. Rev. A},
	volume = {113},
	issue = {1},
	pages = {012617},
	numpages = {6},
	year = {2026},
	month = {Jan},
	publisher = {American Physical Society},
	doi = {10.1103/jy9l-l8hv},
	url = {https://link.aps.org/doi/10.1103/jy9l-l8hv}
}

@article{PhysRevB.109.235432,
	title = {Three-level {D}icke quantum battery},
	author = {Yang, D.-L. and Yang, F.-M. and Dou, F.-Q.},
	journal = {Phys. Rev. B},
	volume = {109},
	issue = {23},
	pages = {235432},
	numpages = {12},
	year = {2024},
	month = {Jun},
	publisher = {American Physical Society},
	doi = {10.1103/PhysRevB.109.235432},
	url = {https://link.aps.org/doi/10.1103/PhysRevB.109.235432}
}

@article{z8gv-7yyk,
	title = {Role of Matter Interactions in Superradiant Phenomena},
	author = {Mendon\ifmmode \mbox{\c{c}}\else \c{c}\fi{}a, Jo Pedro and Jachymski, Krzysztof and Wang, Yao},
	journal = {Phys. Rev. Lett.},
	volume = {135},
	issue = {13},
	pages = {133601},
	numpages = {7},
	year = {2025},
	month = {Sep},
	publisher = {American Physical Society},
	doi = {10.1103/z8gv-7yyk},
	url = {https://link.aps.org/doi/10.1103/z8gv-7yyk}
}

@article{PhysRevE.87.042123,
	title = {Entanglement boost for extractable work from ensembles of quantum batteries},
	author = {Alicki, Robert and Fannes, Mark},
	journal = {Phys. Rev. E},
	volume = {87},
	issue = {4},
	pages = {042123},
	numpages = {4},
	year = {2013},
	month = {Apr},
	publisher = {American Physical Society},
	doi = {10.1103/PhysRevE.87.042123},
	url = {https://link.aps.org/doi/10.1103/PhysRevE.87.042123}
}

@article{Kosloff2013,
	title = {Quantum Thermodynamics: A Dynamical Viewpoint},
	volume = {15},
	ISSN = {1099-4300},
	url = {http://dx.doi.org/10.3390/e15062100},
	DOI = {10.3390/e15062100},
	number = {6},
	journal = {Entropy},
	publisher = {MDPI AG},
	author = {Kosloff,  Ronnie},
	year = {2013},
	month = may,
	pages = {2100–2128}
}

@article{Goold2016,
	title = {The role of quantum information in thermodynamics—a topical review},
	volume = {49},
	ISSN = {1751-8121},
	url = {http://dx.doi.org/10.1088/1751-8113/49/14/143001},
	DOI = {10.1088/1751-8113/49/14/143001},
	number = {14},
	journal = {J. Phys. A: Math. Theor.},
	publisher = {IOP Publishing},
	author = {Goold,  John and Huber,  Marcus and Riera,  Arnau and Rio,  Lídia del and Skrzypczyk,  Paul},
	year = {2016},
	month = {Feb},
	pages = {143001}
}

@article{PhysRev.93.99,
	title = {Coherence in Spontaneous Radiation Processes},
	author = {Dicke, R. H.},
	journal = {Phys. Rev.},
	volume = {93},
	issue = {1},
	pages = {99--110},
	numpages = {0},
	year = {1954},
	month = {Jan},
	publisher = {American Physical Society},
	doi = {10.1103/PhysRev.93.99},
	url = {https://link.aps.org/doi/10.1103/PhysRev.93.99}
}

@article{Gross1982,
	title = {Superradiance: An essay on the theory of collective spontaneous emission},
	volume = {93},
	ISSN = {0370-1573},
	url = {http://dx.doi.org/10.1016/0370-1573(82)90102-8},
	DOI = {10.1016/0370-1573(82)90102-8},
	number = {5},
	journal = {Phys. Rep.},
	publisher = {Elsevier BV},
	author = {Gross,  M. and Haroche,  S.},
	year = {1982},
	month = dec,
	pages = {301–396}
}

@article{PhysRevA.102.023718,
	title = {Gauge invariance of the {Dicke} and {Hopfield} models},
	author = {Garziano, Luigi and Settineri, Alessio and Di Stefano, Omar and Savasta, Salvatore and Nori, Franco},
	journal = {Phys. Rev. A},
	volume = {102},
	issue = {2},
	pages = {023718},
	numpages = {11},
	year = {2020},
	month = {Aug},
	publisher = {American Physical Society},
	doi = {10.1103/PhysRevA.102.023718},
	url = {https://link.aps.org/doi/10.1103/PhysRevA.102.023718}
}

@article{PhysRevResearch.6.033181,
	title = {{Lee-Yang} theory of the superradiant phase transition in the open {Dicke} model},
	author = {Brange, Fredrik and Lambert, Neill and Nori, Franco and Flindt, Christian},
	journal = {Phys. Rev. Res.},
	volume = {6},
	issue = {3},
	pages = {033181},
	numpages = {10},
	year = {2024},
	month = {Aug},
	publisher = {American Physical Society},
	doi = {10.1103/PhysRevResearch.6.033181},
	url = {https://link.aps.org/doi/10.1103/PhysRevResearch.6.033181}
}

@article{PhysRevLett.117.073003,
	title = {Observation of Single-Photon Superradiance and the Cooperative {Lamb} Shift in an Extended Sample of Cold Atoms},
	author = {Roof, S. J. and Kemp, K. J. and Havey, M. D. and Sokolov, I. M.},
	journal = {Phys. Rev. Lett.},
	volume = {117},
	issue = {7},
	pages = {073003},
	numpages = {5},
	year = {2016},
	month = {Aug},
	publisher = {American Physical Society},
	doi = {10.1103/PhysRevLett.117.073003},
	url = {https://link.aps.org/doi/10.1103/PhysRevLett.117.073003}
}

@article{Garraway2011,
	title = {The {Dicke} model in quantum optics: {Dicke} model revisited},
	volume = {369},
	ISSN = {1471-2962},
	url = {http://dx.doi.org/10.1098/rsta.2010.0333},
	DOI = {10.1098/rsta.2010.0333},
	number = {1939},
	journal = {Phil. Trans. R. Soc. A},
	publisher = {The Royal Society},
	author = {Garraway,  Barry M.},
	year = {2011},
	month = mar,
	pages = {1137–1155}
}

@book{Marlan1997,
	author    = {Scully,  Marlan O. and Zubairy,  M. Suhail},
	title     = {Quantum Optics},
	edition   = {1},
	publisher = {Cambridge University Press},
	year      = {1997}
}

@article{PhysRevE.67.066203,
	title = {Chaos and the quantum phase transition in the {Dicke} model},
	author = {Emary, Clive and Brandes, Tobias},
	journal = {Phys. Rev. E},
	volume = {67},
	issue = {6},
	pages = {066203},
	numpages = {22},
	year = {2003},
	month = {Jun},
	publisher = {American Physical Society},
	doi = {10.1103/PhysRevE.67.066203},
	url = {https://link.aps.org/doi/10.1103/PhysRevE.67.066203}
}

@book{Agarwal2012,
	title = {Quantum Optics},
	ISBN = {9781139035170},
	url = {http://dx.doi.org/10.1017/CBO9781139035170},
	DOI = {10.1017/cbo9781139035170},
	publisher = {Cambridge University Press},
	author = {Agarwal,  Girish S.},
	year = {2012},
	month = nov 
}

@article{Kirton2019,
	title = {Introduction to the {Dicke} Model: From Equilibrium to Nonequilibrium,  and Vice Versa},
	volume = {2},
	ISSN = {2511-9044},
	number = {1–2},
	journal = {Adv. Quantum Technol.},
	publisher = {Wiley},
	author = {Kirton,  Peter and Roses,  Mor M. and Keeling,  Jonathan and Dalla Torre,  Emanuele G.},
	year = {2019},
	url = {http://dx.doi.org/10.1002/qute.201970013},
	month = {Feb} 

}

@article{PhysRev.132.2521,
	title = {Ferromagnetism and Spin Waves in the Band Theory},
	author = {Mattis, Daniel C.},
	journal = {Phys. Rev.},
	volume = {132},
	issue = {6},
	pages = {2521--2528},
	numpages = {0},
	year = {1963},
	month = {Dec},
	publisher = {American Physical Society},
	doi = {10.1103/PhysRev.132.2521},
	url = {https://link.aps.org/doi/10.1103/PhysRev.132.2521}
}

@article{PhysRevResearch.5.033002,
	title = {Generalized Dicke model and gauge-invariant master equations for two atoms in ultrastrongly-coupled cavity quantum electrodynamics},
	author = {Akbari, Kamran and Salmon, Will and Nori, Franco and Hughes, Stephen},
	journal = {Phys. Rev. Res.},
	volume = {5},
	issue = {3},
	pages = {033002},
	numpages = {16},
	year = {2023},
	month = {Jul},
	publisher = {American Physical Society},
	doi = {10.1103/PhysRevResearch.5.033002},
	url = {https://link.aps.org/doi/10.1103/PhysRevResearch.5.033002}
}

@article{Ma2011,
	title = {Quantum spin squeezing},
	volume = {509},
	ISSN = {0370-1573},
	url = {http://dx.doi.org/10.1016/j.physrep.2011.08.003},
	DOI = {10.1016/j.physrep.2011.08.003},
	number = {2–3},
	journal = {Phys. Rep.},
	publisher = {Elsevier BV},
	author = {Ma,  Jian and Wang,  Xiao Guang and Sun,  C. P. and Nori,  Franco},
	year = {2011},
	month = dec,
	pages = {89–165}
}

@article{PhysRevA.47.5138,
	title = {Squeezed spin states},
	author = {Kitagawa, Masahiro and Ueda, Masahito},
	journal = {Phys. Rev. A},
	volume = {47},
	issue = {6},
	pages = {5138--5143},
	numpages = {0},
	year = {1993},
	month = {Jun},
	publisher = {American Physical Society},
	doi = {10.1103/PhysRevA.47.5138},
	url = {https://link.aps.org/doi/10.1103/PhysRevA.47.5138}
}

@article{PhysRev.58.1098,
	title = {Field Dependence of the Intrinsic Domain Magnetization of a Ferromagnet},
	author = {Holstein, T. and Primakoff, H.},
	journal = {Phys. Rev.},
	volume = {58},
	issue = {12},
	pages = {1098--1113},
	numpages = {0},
	year = {1940},
	month = {Dec},
	publisher = {American Physical Society},
	doi = {10.1103/PhysRev.58.1098},
	url = {https://link.aps.org/doi/10.1103/PhysRev.58.1098}
}

@article{PhysRevA.68.033821,
	title = {Relations between bosonic quadrature squeezing and atomic spin squeezing},
	author = {Wang, Xiaoguang and Sanders, Barry C.},
	journal = {Phys. Rev. A},
	volume = {68},
	issue = {3},
	pages = {033821},
	numpages = {8},
	year = {2003},
	month = {Sep},
	publisher = {American Physical Society},
	doi = {10.1103/PhysRevA.68.033821},
	url = {https://link.aps.org/doi/10.1103/PhysRevA.68.033821}
}

@misc{YKX260518,
	howpublished = {\url{https://link.aps.org/supplemental/XXXXXXX}},
	note         = {See {Supplemental Material} for a detailed derivation and possible experimiental implementation.}
}

@article{PhysRevA.6.2211,
	title = {Atomic Coherent States in Quantum Optics},
	author = {Arecchi, F. T. and Courtens, Eric and Gilmore, Robert and Thomas, Harry},
	journal = {Phys. Rev. A},
	volume = {6},
	issue = {6},
	pages = {2211--2237},
	numpages = {0},
	year = {1972},
	month = {Dec},
	publisher = {American Physical Society},
	doi = {10.1103/PhysRevA.6.2211},
	url = {https://link.aps.org/doi/10.1103/PhysRevA.6.2211}
}

@article{Hu2022,
	title = {Optimal charging of a superconducting quantum battery},
	volume = {7},
	ISSN = {2058-9565},
	url = {http://dx.doi.org/10.1088/2058-9565/ac8444},
	DOI = {10.1088/2058-9565/ac8444},
	number = {4},
	journal = {Quantum Sci. and Technol.},
	publisher = {IOP Publishing},
	author = {Hu,  Chang-Kang and Qiu,  Jiawei and Souza,  Paulo J P and Yuan,  Jiahao and Zhou,  Yuxuan and Zhang,  Libo and Chu,  Ji and Pan,  Xianchuang and Hu,  Ling and Li,  Jian and Xu,  Yuan and Zhong,  Youpeng and Liu,  Song and Yan,  Fei and Tan,  Dian and Bachelard,  R and Villas-Boas,  C J and Santos,  Alan C and Yu,  Dapeng},
	year = {2022},
	month = Aug,
	pages = {045018}
}

@article{Camposeo2025,
	title = {Quantum Batteries: A Materials Science Perspective},
	volume = {37},
	ISSN = {1521-4095},
	url = {http://dx.doi.org/10.1002/adma.202415073},
	number = {17},
	journal = {Adv. Mater.},
	publisher = {Wiley},
	author = {Camposeo,  Andrea and Virgili,  Tersilla and Lombardi,  Floriana and Cerullo,  Giulio and Pisignano,  Dario and Polini,  Marco},
	year = {2025},
	month = {Feb} 
}

@book{Carmichael1993,
	title = {An Open Systems Approach to Quantum Optics: Lectures Presented at the Université Libre de Bruxelles October 28 to November 4,  1991},
	ISBN = {9783540476207},
	ISSN = {0940-7677},
	url = {http://dx.doi.org/10.1007/978-3-540-47620-7},
	DOI = {10.1007/978-3-540-47620-7},
	journal = {Lecture Notes in Physics Monographs},
	publisher = {Springer Berlin Heidelberg},
	author = {Carmichael,  Howard},
	year = {1993}
}

@book{Walls2025,
	title = {Quantum Optics},
	ISBN = {9783031841774},
	ISSN = {1868-4521},
	url = {http://dx.doi.org/10.1007/978-3-031-84177-4},
	DOI = {10.1007/978-3-031-84177-4},
	journal = {Graduate Texts in Physics},
	publisher = {Springer Nature Switzerland},
	author = {Walls,  D. F. and Milburn,  Gerard J.},
	year = {2025}
}

@article{PhysRevB.98.205423,
	title = {Charger-mediated energy transfer in exactly solvable models for quantum batteries},
	author = {Andolina, Gian Marcello and Farina, Donato and Mari, Andrea and Pellegrini, Vittorio and Giovannetti, Vittorio and Polini, Marco},
	journal = {Phys. Rev. B},
	volume = {98},
	issue = {20},
	pages = {205423},
	numpages = {11},
	year = {2018},
	month = {Nov},
	publisher = {American Physical Society},
	doi = {10.1103/PhysRevB.98.205423},
	url = {https://link.aps.org/doi/10.1103/PhysRevB.98.205423}
}

@article{PhysRevA.100.022115,
	title = {Thermalization of a {Lipkin-Meshkov-Glick} model coupled to a bosonic bath},
	author = {Louw, Jan C. and Kriel, Johannes N. and Kastner, Michael},
	journal = {Phys. Rev. A},
	volume = {100},
	issue = {2},
	pages = {022115},
	numpages = {9},
	year = {2019},
	month = {Aug},
	publisher = {American Physical Society},
	doi = {10.1103/PhysRevA.100.022115},
	url = {https://link.aps.org/doi/10.1103/PhysRevA.100.022115}
}

@article{PhysRevE.79.041129,
	title = {Quantum thermodynamic cycles and quantum heat engines},
	author = {Quan, H. T.},
	journal = {Phys. Rev. E},
	volume = {79},
	issue = {4},
	pages = {041129},
	numpages = {10},
	year = {2009},
	month = {Apr},
	publisher = {American Physical Society},
	doi = {10.1103/PhysRevE.79.041129},
	url = {https://link.aps.org/doi/10.1103/PhysRevE.79.041129}
}

@article{PhysRevLett.120.093601,
	title = {Exponentially Enhanced Light-Matter Interaction, Cooperativities, and Steady-State Entanglement Using Parametric Amplification},
	author = {Qin, Wei and Miranowicz, Adam and Li, Peng-Bo and L\"u, Xin-You and You, J. Q. and Nori, Franco},
	journal = {Phys. Rev. Lett.},
	volume = {120},
	issue = {9},
	pages = {093601},
	numpages = {7},
	year = {2018},
	month = {Mar},
	publisher = {American Physical Society},
	doi = {10.1103/PhysRevLett.120.093601},
	url = {https://link.aps.org/doi/10.1103/PhysRevLett.120.093601}
}

@article{Qin2024,
	title = {Quantum amplification and simulation of strong and ultrastrong coupling of light and matter},
	volume = {1078},
	ISSN = {0370-1573},
	url = {http://dx.doi.org/10.1016/j.physrep.2024.05.003},
	DOI = {10.1016/j.physrep.2024.05.003},
	journal = {Phys. Rep.},
	publisher = {Elsevier BV},
	author = {Qin,  Wei and Kockum,  Anton Frisk and Muñoz,  Carlos Sánchez and Miranowicz,  Adam and Nori,  Franco},
	year = {2024},
	month = Aug,
	pages = {1–59}
}

@article{Ashhab2006,
	title = {Rabi oscillations in a qubit coupled to a quantum two-level system},
	volume = {8},
	ISSN = {1367-2630},
	url = {http://dx.doi.org/10.1088/1367-2630/8/6/103},
	DOI = {10.1088/1367-2630/8/6/103},
	number = {6},
	journal = {New J. Phys.},
	publisher = {IOP Publishing},
	author = {Ashhab,  S and Johansson,  J R and Nori,  Franco},
	year = {2006},
	month = {June},
	pages = {103–103}
}

@article{PhysRevB.78.054512,
	title = {Pseudo-{Rabi} oscillations in superconducting flux qubits in the classical regime},
	author = {Omelyanchouk, A. N. and Shevchenko, S. N. and Zagoskin, A. M. and Il'ichev, E. and Nori, Franco},
	journal = {Phys. Rev. B},
	volume = {78},
	issue = {5},
	pages = {054512},
	numpages = {5},
	year = {2008},
	month = {Aug},
	publisher = {American Physical Society},
	doi = {10.1103/PhysRevB.78.054512},
	url = {https://link.aps.org/doi/10.1103/PhysRevB.78.054512}
}

@article{PhysRevB.80.212503,
	title = {Noise-induced quantum coherence and persistent {Rabi} oscillations in a {Josephson} flux qubit},
	author = {Omelyanchouk, A. N. and Savel'ev, S. and Zagoskin, A. M. and Il'ichev, E. and Nori, Franco},
	journal = {Phys. Rev. B},
	volume = {80},
	issue = {21},
	pages = {212503},
	numpages = {4},
	year = {2009},
	month = {Dec},
	publisher = {American Physical Society},
	doi = {10.1103/PhysRevB.80.212503},
	url = {https://link.aps.org/doi/10.1103/PhysRevB.80.212503}
}

@article{PhysRevA.92.063830,
	title = {Multiphoton quantum {Rabi} oscillations in ultrastrong cavity {QED}},
	author = {Garziano, Luigi and Stassi, Roberto and Macr\`{\i}, Vincenzo and Kockum, Anton Frisk and Savasta, Salvatore and Nori, Franco},
	journal = {Phys. Rev. A},
	volume = {92},
	issue = {6},
	pages = {063830},
	numpages = {10},
	year = {2015},
	month = {Dec},
	publisher = {American Physical Society},
	doi = {10.1103/PhysRevA.92.063830},
	url = {https://link.aps.org/doi/10.1103/PhysRevA.92.063830}
}

@article{PhysRevLett.126.023602,
	title = {Shortcuts to Adiabaticity for the Quantum {Rabi} Model: Efficient Generation of Giant Entangled Cat States via Parametric Amplification},
	author = {Chen, Ye-Hong and Qin, Wei and Wang, Xin and Miranowicz, Adam and Nori, Franco},
	journal = {Phys. Rev. Lett.},
	volume = {126},
	issue = {2},
	pages = {023602},
	numpages = {8},
	year = {2021},
	month = {Jan},
	publisher = {American Physical Society},
	doi = {10.1103/PhysRevLett.126.023602},
	url = {https://link.aps.org/doi/10.1103/PhysRevLett.126.023602}
}

@article{PhysRevLett.133.033603,
	title = {Error-Tolerant Amplification and Simulation of the Ultrastrong-Coupling Quantum {Rabi} Model},
	author = {Chen, Ye-Hong and Shi, Zhi-Cheng and Nori, Franco and Xia, Yan},
	journal = {Phys. Rev. Lett.},
	volume = {133},
	issue = {3},
	pages = {033603},
	numpages = {7},
	year = {2024},
	month = {Jul},
	publisher = {American Physical Society},
	doi = {10.1103/PhysRevLett.133.033603},
	url = {https://link.aps.org/doi/10.1103/PhysRevLett.133.033603}
}

\end{document}